\pdfoutput=1

\documentclass[11pt,twoside,a4paper,cmspaper,final,collab]{cms-tdr}
\def\svnVersion{182565}\def\cmsCernNo{2012-293}\def\cmsCernDate{\today}\def\cmsMessage{Submitted to Physics Letters B}

\begin{document}\cmsNoteHeader{FWD-11-001}

\hyphenation{had-ron-i-za-tion}
\hyphenation{cal-or-i-me-ter}
\hyphenation{de-vices}
\RCS$Revision: 144475 $
\RCS$HeadURL: svn+ssh://svn.cern.ch/reps/tdr2/papers/FWD-11-001/trunk/FWD-11-001.tex $
\RCS$Id: FWD-11-001.tex 144475 2012-08-20 20:49:03Z nicolo $
\providecommand{\re}{\ensuremath{\mathrm{e}}}
\providecommand{\lum}{\ensuremath{\,(\text{lum.})}\xspace} \newlength\cmsFigWidth
\ifthenelse{\boolean{cms@external}}{\setlength\cmsFigWidth{\columnwidth}}{\setlength\cmsFigWidth{0.8\textwidth}}
\ifthenelse{\boolean{cms@external}}{\providecommand{\cmsLeft}{top}}{\providecommand{\cmsLeft}{left}}
\ifthenelse{\boolean{cms@external}}{\providecommand{\cmsRight}{bottom}}{\providecommand{\cmsRight}{right}}
\cmsNoteHeader{FWD-11-001} % This is over-written in the CMS environment: useful as preprint no. for export versions
\title{Measurement of the inelastic proton-proton cross section at $\sqrt{s} = 7\TeV$}% Force line breaks with \\

\date{\today}

\abstract{
A measurement is presented of the inelastic proton-proton cross section  at a centre-of-mass energy of $\sqrt{s} = 7$\TeV. Using the CMS detector at the LHC, the inelastic cross section is measured
through two independent methods based on information from  (i) forward calorimetry (for pseudorapidity $ 3 < |\eta| < 5$), in collisions where at least one proton  loses  more than $5\times10^{-6}$ of its longitudinal momentum,
and  (ii) the central tracker ( $|\eta| < 2.4$), in collisions containing an interaction vertex with more than one, two, or three tracks  with  transverse momenta $\pt > 200$\MeVc.  The measurements cover a large fraction of the inelastic cross section for
particle production over about nine units of pseudorapidity and down to small transverse momenta. The results are compared with those of other experiments, and with models used to describe high-energy hadronic interactions.}

\hypersetup{%
pdfauthor={CMS Collaboration},%
pdftitle={Measurement of the inelastic proton-proton cross section at sqrt(s) = 7 TeV},%
pdfsubject={CMS},%
pdfkeywords={CMS, physics, inelastic cross section, tracking efficiency, forward energy, total cross section, LHC}}

\maketitle %maketitle comes after all the front information has been supplied

\section{Introduction}
\label{Introduction}

Total hadronic cross sections, as well as their major subdivisions into
elastic, inelastic diffractive  and inelastic non-diffractive contributions, comprise fundamental quantities that have
been studied in high-energy particle, nuclear, and cosmic-ray physics over the past 60 years, in experiments covering many orders of
magnitude in centre-of-mass energy~\cite{PhysRevD.86.010001,Antchev:2011vsR, ATLASXSEC, PierreAuger:2011aaR, :2012sj}.

The bulk of the total cross section in proton-proton (pp)  hadronic interactions cannot be calculated through perturbative  quantum
chromodynamics,  but  phenomenological approaches based on
fundamental principles of quantum mechanics, such as unitarity and analyticity, can be used to accommodate the experimental results (\eg Ref.~\cite{Donnachie:2002en}, and references therein). Although  phenomenological models of cross sections at  low centre-of-mass energies ($\sqrt{s}\le 100$\GeV) provide a rather precise description of the data, there are large uncertainties in extrapolating to the energy range of the Large Hadron Collider (LHC). The  measured  inelastic pp cross section ($\sigma_{\text{inel}}$) serves as an input to these
phenomenological models, and provides basic information needed for tuning hadronic Monte Carlo (MC) generators.
The values of $\sigma_{\text{inel}}$ are also used to estimate the number of pp interactions as a function of luminosity at colliders, and are relevant to studies of high-energy cosmic rays~\cite{UlrichR} and to the  characterization of  global properties of heavy-ion collisions, especially  in the context of the Glauber model~\cite{GlauberR}.

This Letter presents a measurement of the inelastic pp cross
section at $\sqrt{s}=7\TeV$, using data collected with the Compact Muon Solenoid (CMS) detector at the LHC.
The analysis is based mostly on the central silicon tracker and the forward hadron calorimeters (HF)
of the CMS apparatus. The combination of these two detectors provides sensitivity to a large part of the inelastic cross
section, including central diffractive production, where particles can be produced at small values of  pseudorapidity.

The measurement using the HF calorimeters covers a region  of phase space corresponding to values of
fractional momentum loss of the scattered proton of $\xi =
(M_Xc^2)^2/s > 5\times 10^{-6}$, equivalent to $M_X > 16$\GeVcc, where  $M_X$ is defined as the larger mass of the two dissociated proton systems in the final state. This coverage is  the same as that used in recent publications by the ATLAS~\cite{ATLASXSEC} and the ALICE~\cite{:2012sj} Collaborations.

\section{Experimental apparatus}

\label{Detector}

A detailed description of the CMS apparatus can be found in Ref.~\cite{:2008zzk},  and the features most relevant  to the present analysis are sketched below.
The CMS detector comprises a 6\unit{m} diameter, 13\unit{m}
long,  3.8\unit{T} solenoid magnet, with a combined silicon pixel and strip tracker
covering the region $|\eta| <2.5$, a lead-tungstate electromagnetic calorimeter  and a brass/scintillator hadronic calorimeter covering the region $|\eta| <$~3.0; these detectors are contained within the volume of the magnetic field. The pseudorapidity is defined as $\eta = -\ln\left[\tan\left(\theta/2\right)\right]$, where $\theta$ is the polar angle of any particle with respect to the anticlockwise circulating beam.  Several layers of muon chambers (drift-tube, resistive-plate  and cathode-strip chambers) form the outer part of the detector. The charged-particle resolution of the central tracker for a transverse momentum   of $\pt=1\GeVc$ is between 0.7\% at  $\abs{\eta}=0$ and 2\% at  $\abs{\eta}=2.5$ \cite{:2008zzk}.

On each side of the detector, at $3.0 < |\eta| < 5.2$, reside the hadron forward calorimeters (HF),  each  composed of 18 iron wedges, with embedded quartz fibres running along the beam direction. Each wedge is
subdivided into 13 $\eta$-segments, called towers.

The beam-sensitive ``pick-up'' detectors, consisting of two pairs of button
electrodes located  at ${\pm}175$\unit{m} from the centre of the detector,  provide almost 100\%
detection efficiency and accurate timing of proton bunches at CMS. The luminosity is calculated from dedicated Van der Meer scans, using information from the beam profile and beam current measurements, with a precision of 4\% that is dominated by the uncertainty of the beam current determination~\cite{:vdmeer,:CMSlumi}.

\section{Estimating the inelastic cross section using the HF calorimeters}
\label{HFMethod}
In this method, the inelastic pp cross section is measured by counting the number of  events that deposit at least 5\GeV of energy in either of the two HF calorimeters.
 The threshold $E_{\mathrm{HF}} > 5$\GeV is set to minimize the effect of detector noise on the  efficiency of selecting pp collisions.

\subsection{Event selection and analysis}

The analysis is performed using data collected in low-luminosity runs with an average of 0.007 to 0.11  collisions per bunch crossing.  The events are collected using three triggers: (i) a coincidence trigger that requires the presence of two colliding bunches,  used to select an unbiased sample of pp events, (ii) a single-bunch trigger, requiring the presence of just one unpaired bunch, used to estimate beam-induced backgrounds, and (iii) a random ``empty'' trigger, requiring  absence of both beams, which is used to estimate detector noise. All these triggers are formed from information  provided by the beam pick-up detectors.

The analysis is based on counting the number of pp collisions  with  $ E_{\mathrm{HF}} > 5$\GeV in either of the two HF calorimeters. The cross section is evaluated in terms of the variable $\xi$, which is defined through MC studies as follows.
For each MC event, generator-level information is used to order final-state particles in rapidity  and to find the
largest gap  between two consecutive particles. This ``central'' gap  is used to separate all particles into two groups, by assigning each
particle, according to its  rapidity position relative to that gap, to system $A$ or system $B$. Finally, the masses of system $A$ and $B$ are calculated, and the larger of the two is called  $M_X$, while the smaller one $M_Y$, thereby  defining $\xi = (M_Xc^2)^2/s$. In single-diffractive events,  $\xi$ corresponds to  the fraction of momentum lost by the proton in the collision. The $\xi$ distribution is bound by the elastic limit of $\log_{10}((m_\text{proton}c^2)^2/s) \approx -7.75$.

\begin{figure*}[htbp]
  \begin{center}
\includegraphics[width=0.48\textwidth]{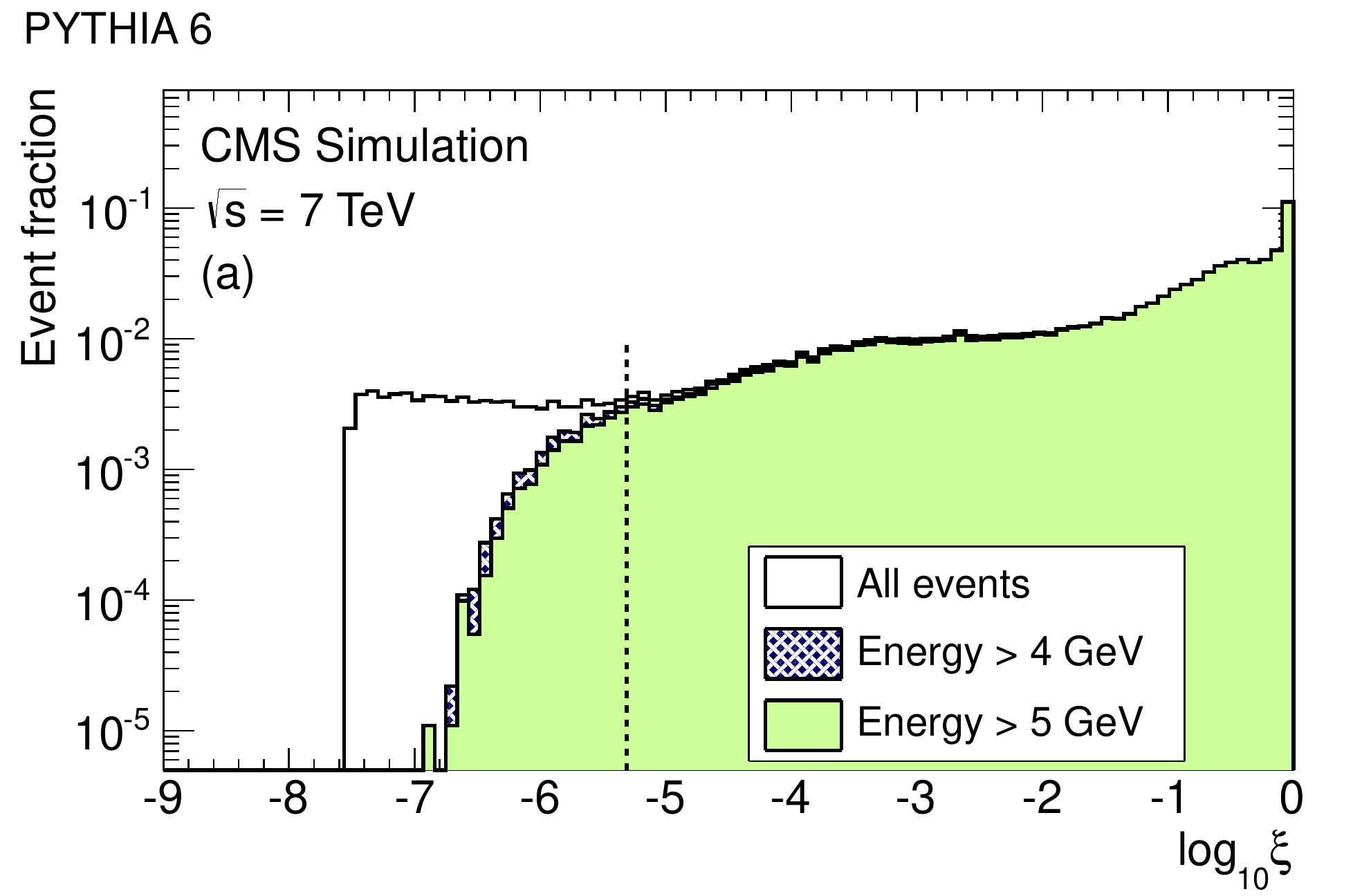}
\includegraphics[width=0.48\textwidth]{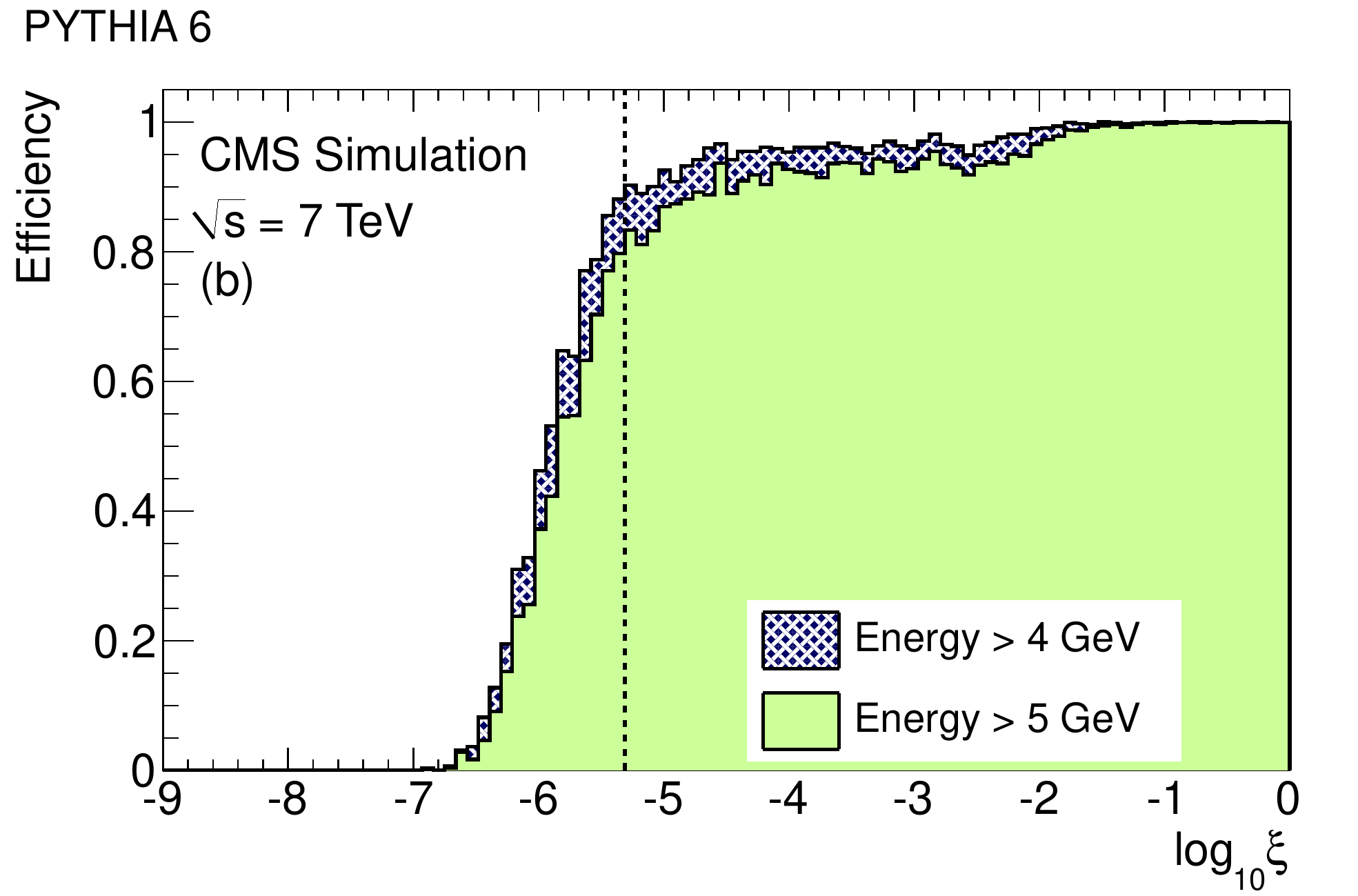}

\includegraphics[width=0.48\textwidth]{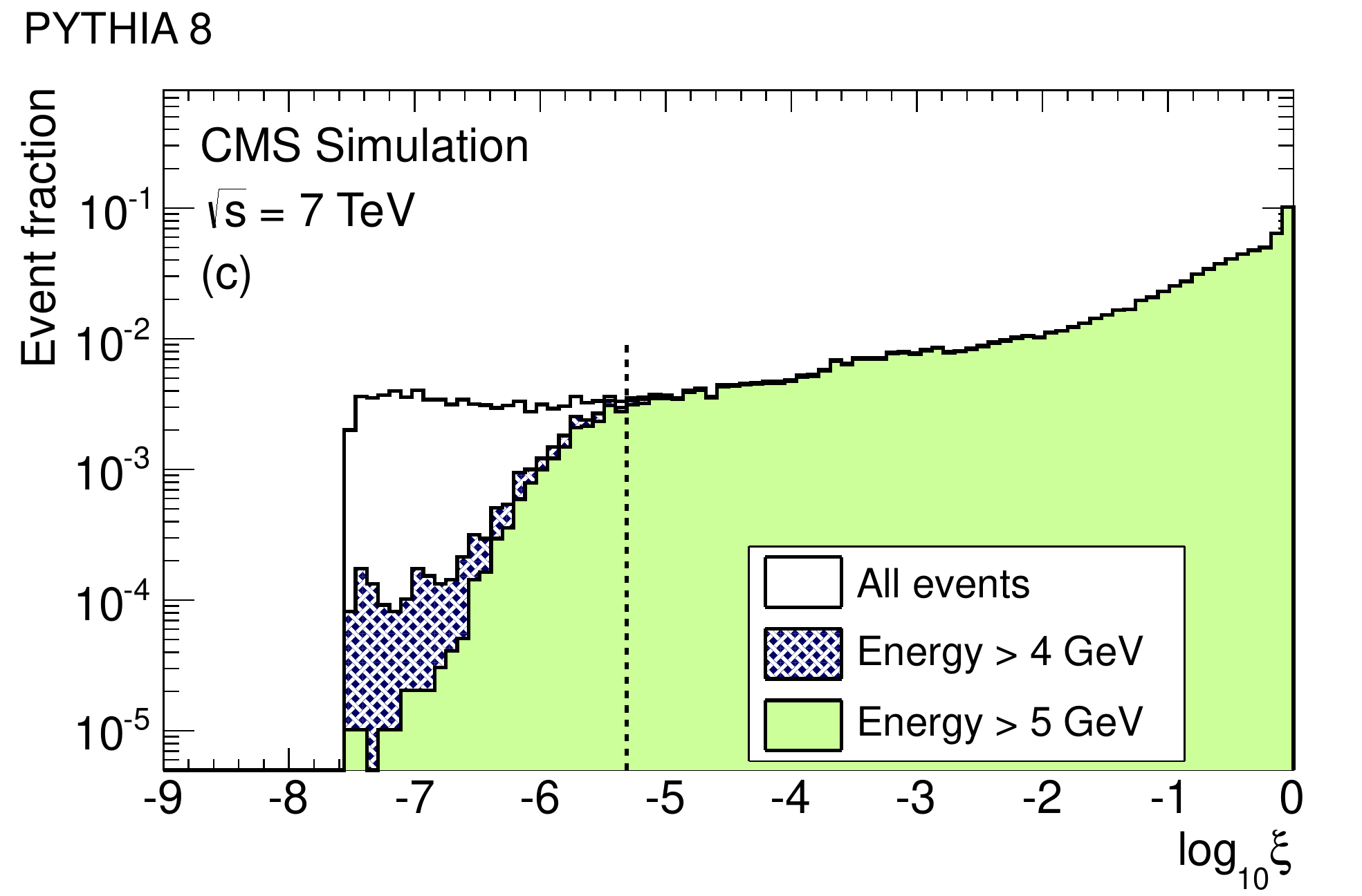}
\includegraphics[width=0.48\textwidth]{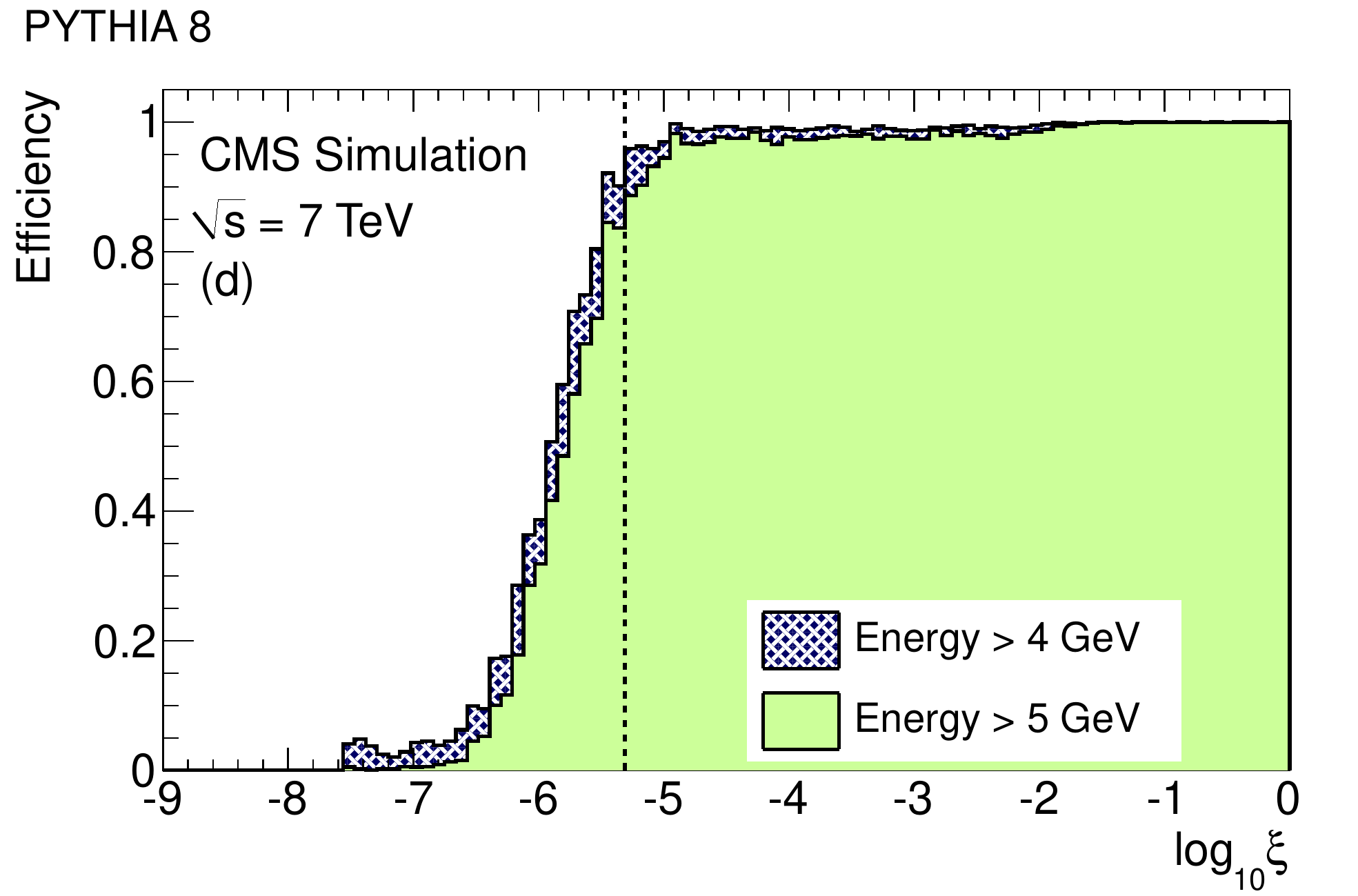}

\includegraphics[width=0.48\textwidth]{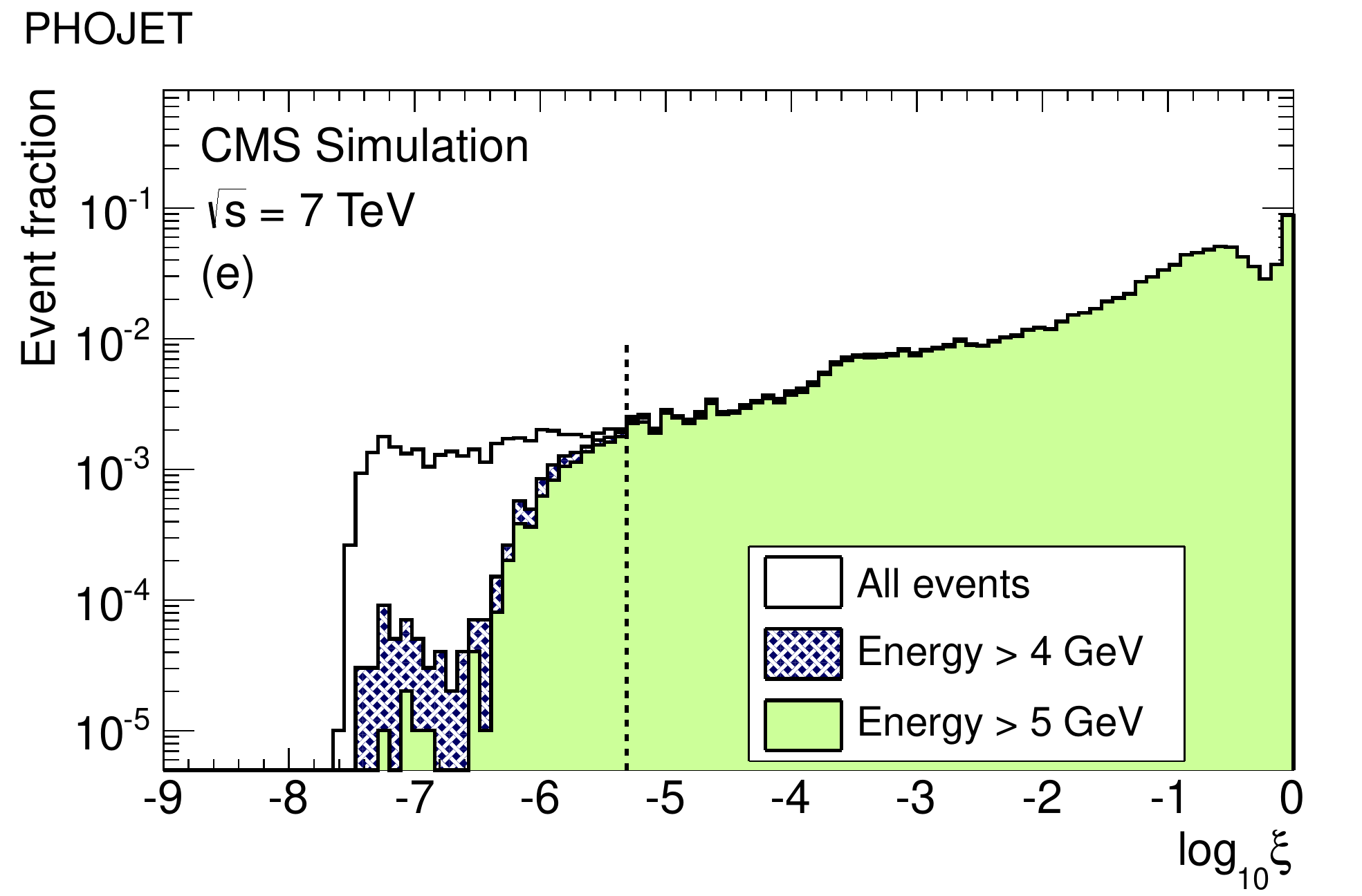}
\includegraphics[width=0.48\textwidth]{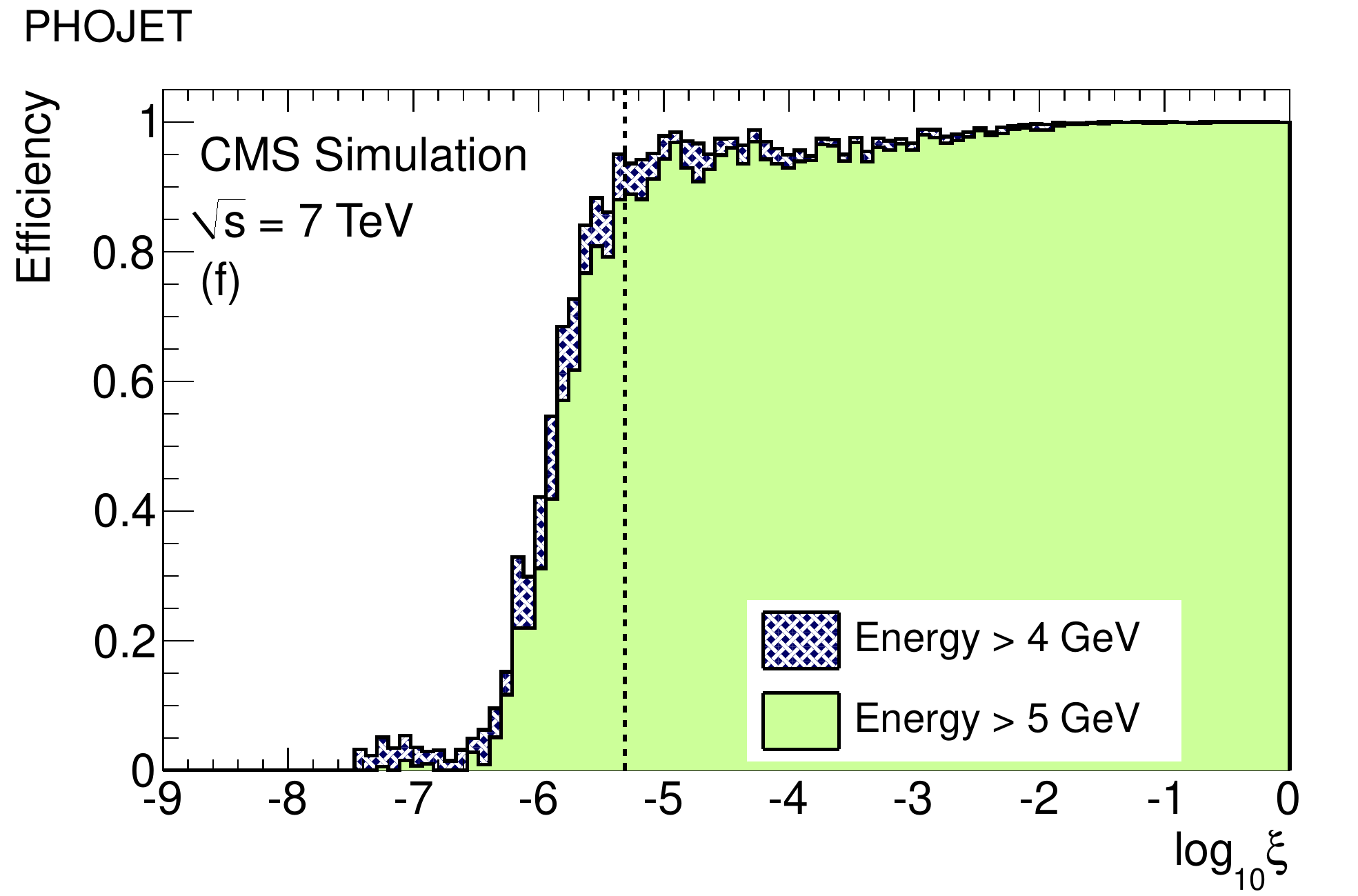}

    \caption{The  normalized $\xi$ distributions  for $E_{\mathrm{HF}} > 4$ and $E_{\mathrm{HF}} > 5$\GeV from MC simulation of inelastic pp collisions using (a) \PYTHIA6, (c) \PYTHIA8, and (e) \textsc{ Phojet}, are shown  for the full range of $\xi$. The  corresponding   efficiencies are shown in (b), (d), and (f), respectively. The cut value of $\xi$ used in this analysis of  $5\times10^{-6}$ is shown on the plots as a dashed vertical line. }
    \label{fig:MxDistribution}
  \end{center}
\end{figure*}

The distributions in $\xi$ values for $E_{\mathrm{HF}}>4$ and $>5$\GeV
 are shown in  Fig.~\ref{fig:MxDistribution} for three Monte Carlo models: \PYTHIA6 (version 6.422)~\cite{Pythia6R}, \PYTHIA8 (version 8.135, 8.145)~\cite{ Pythia8R}, and \textsc{Phojet} (version 1.12-35)~\cite{PhojetR,:PHOJET}. These selected models differ in the treatment of non-perturbative processes and use a different set of assumptions for soft pp interactions. They  capture qualitative features of diffraction well, and they  also cover  reasonable  variations of simulated distributions of $\xi$.  As the plots illustrate, to maintain large detection efficiency, and to mitigate model-dependence, it appears adequate to restrict the range of $\xi$ to values greater than $ 5\times10^{-6}$.
The measured values of $\sigma_{\text{inel}}$  are corrected using  two quantities obtained through MC simulation:  the selection efficiency $\epsilon_{\xi}$, which represents the fraction of pp interactions with $\xi > 5\times10^{-6}$ that are selected by requiring  $E_{\mathrm{HF}} > 5$\GeV, and the contamination $b_{\xi}$, which is the fraction of events that have  $E_{\mathrm{HF}} > 5$\GeV, but originate from  $\xi < 5\times10^{-6}$. Table~\ref{tab:ksiEfficency} gives the values of  $\epsilon_{\xi}$  and  $b_{\xi}$  estimated in the three Monte Carlo models. These efficiencies carry a small ($<1\%$) uncertainty due to the HF energy scale uncertainty, estimated as the difference between the efficiencies obtained with different HF energy thresholds (corresponding to 20\% energy scale variations).
As the table shows, the criterion $E_{\mathrm{HF}} > 5$\GeV selects a large fraction of events with $\xi > 5\times10^{-6}$, with only a small contamination from events with  $\xi < 5\times10^{-6}$ that characterize contributions originating from low-mass single-proton or double-proton fragmentation.

\begin{table}[htb]
  \begin{center}
\topcaption{Values of efficiency ($\epsilon_\xi$) and contamination ($b_\xi$) for  events with $\xi > 5\times10^{-6}$ using the selection criterion of $E_{\mathrm{HF}} > 5$\GeV, obtained for three Monte Carlo models of hadronic production.}
  \begin{tabular}{l  c c } \hline
Generator  & $\epsilon_{\xi}$ (\%) & $b_{\xi}$ (\%) \\ \hline
\PYTHIA6 & $97.5 \pm 0.6$   & 2.0 \\
\PYTHIA8 & $99.3 \pm 0.2$   & 2.0 \\
\textsc{phojet}  & $99.1 \pm 0.2$   & 1.2 \\
\hline
\end{tabular}
\label{tab:ksiEfficency}
\end{center}
\end{table}

\subsection{Measurement of the inelastic cross section}
\label{Method}

The analysis is performed using $\approx$9.2 million  events,  corresponding to an integrated luminosity of 2.78\mubinv, collected under the two-bunch coincidence condition, of which  2.1\% have $E_{\mathrm{HF}} > 5$\GeV.
The fractions of  $E_{\mathrm{HF}} > 5$\GeV events  selected by the single-bunch and empty triggers are, respectively, 0.30\% and 0.32\%, suggesting that most of the single-bunch events are from  detector noise rather than beam-gas collisions. This is confirmed by the observation  that, in the single-bunch triggered sample, the number of events with at least one track is very small.  For this reason, beam-gas contributions are considered negligible.

The number of detected inelastic collisions  ($N_{\text{inel}}$) contained in the total number of coincidence trigger  events  ($ N_{\text{coinc}}$) is obtained as follows:
\begin{equation}
\label{eq:Corrected}
N_{\text{inel}} = N_{\text{coinc}}[(\mathcal{F}_{\text{coinc}} - \mathcal{F}_{\text{empty}}) + \mathcal{F}_{\text{empty}} (\mathcal{F}_{\text{coinc}} - \mathcal{F}_{\text{empty}})],
\end{equation}
where $ \mathcal{F}_{\text{empty}}$ and $ \mathcal{F}_{\text{coinc}}$ correspond to the fractions of empty and coincidence triggers with  $E_{\mathrm{HF}} > 5$\GeV.
The term $ N_{\text{coinc}}\mathcal{F}_{\text{empty}} (\mathcal{F}_{\text{coinc}} - \mathcal{F}_{\text{empty}})$ represents the number of
 true collisions in  $N_{\text{coinc}} \mathcal{F}_{\text{empty}}$  events.

The value of  $N_{\text{inel}}$ has to be corrected for event pileup, i.e. the possibility that more than one collision with $E_\mathrm{HF} > 5$\GeV occurs in the same  trigger, but all such collisions are counted as just a single event. The number of collisions per  trigger is assumed to follow Poisson statistics,  for which the probability of $i$  simultaneous collisions ($i = 1, 2, 3,\ldots$)  is given by
\begin{equation}
\label{eq:Poisson}
P(n,\lambda) = \frac{\lambda^{n} \re^{-\lambda}}{n!},
\end{equation}
where $\lambda$ is the mean number of interactions with $E_\mathrm{HF} > 5$\GeV, which depends on the instantaneous luminosity ($L$).  The fraction $f_{\mathrm{pu}}$ of overlapping collisions, each with $E_\mathrm{HF} > 5$\GeV,  is computed as
\begin{equation}
\label{eq:PUCor}
f_{\mathrm{pu}} = \frac{\Sigma_{n=2}^{\infty} P(n,\lambda)}{\Sigma_{n=1}^{\infty}  P(n,\lambda)}  =  \frac{1 - (1+\lambda)\, \re^{-\lambda}} {1 - \re^{-\lambda}} \sim \frac{\lambda}{2} - \frac{\lambda^2}{12} +\mathcal{O}\left( \lambda^3 \right) ,
\end{equation}
where  $\lambda$ is evaluated from the fraction of detected interactions  $r_\text{int} = N_{\text{inel}} /N_{\text{coinc}}$:
\begin{equation}
\label{eq:PUCorA}
\begin{split}
 r_\text{int} &= \Sigma_{n=1}^{\infty}  P(n,\lambda) =  1 - P(0, \lambda) = 1 - \re^{-\lambda},\\
\lambda  &= - \ln ( 1 -  r_\text{int}).
\end{split}
\end{equation}

The denominator in Eq.~(\ref{eq:PUCor}) assumes independent probabilities for detecting each of the simultaneous collisions, which is a good approximation for $E_{\mathrm{HF}} > 5$\GeV.

Table~\ref{pileupcorr} lists the values of $\lambda$  and $f_{\mathrm{pu}}$, as calculated using the exact formula in Eq.~(\ref{eq:PUCor}), and their statistical uncertainties for different data runs. The accuracy on the  correction factor $f_{\mathrm{pu}}$ is limited mostly by the  number of events in each run.

\begin{table}[htb]
\begin{center}
\topcaption{Mean number of collisions with  $E_{\mathrm{HF}} > 5$\GeV per coincidence trigger ($\lambda$) and fraction of overlapping collisions ($f_{\mathrm{pu}}$) for the runs  used in this analysis.}
\begin{tabular}{l c c }
\hline
Run No. & $\lambda$ & $f_{\mathrm{pu}}$\\
\hline
132601 &  ($0.64 \pm 0.01$) \% & $0.0032 \pm 0.0001$ \\
132599 &  ($0.78 \pm 0.01$) \% & $0.0039 \pm 0.0001$ \\
133877 &  ($1.74 \pm 0.02$) \% & $0.0087 \pm  0.0001$ \\
133874 &  ($3.34 \pm 0.05$) \% & $0.0166 \pm  0.0002$ \\
137027 &  ($4.59 \pm 0.17$) \% & $0.0228 \pm  0.0009$ \\
135575 &  ($8.41 \pm 0.04$) \% & $0.0415 \pm  0.0002$ \\
135175 &  ($9.98 \pm 0.05$) \% & $0.0491 \pm 0.0003$ \\
\hline
\end{tabular}
\label{pileupcorr}
\end{center}
\end{table}

The relationship used to evaluate the cross section for $\xi > 5\times10^{-6}$,
taking  account of corrections for pileup, efficiency, and contamination corresponds to:
\begin{equation}
\label{eq:AccurateCrossSection}
\sigma_{\text{inel}}(\xi > 5\times10^{-6}) =
\frac{N_{\text{inel}}(1-b_{\xi})
(1+f_{\mathrm{pu}})}{\epsilon_{\xi} \int L\, \rd t},
\end{equation}
where $\int L\,\rd t$ is the integrated luminosity of the data sample.

\subsection{Results and systematic uncertainties}
\label{HF_Results}

The value of $\sigma_{\text{inel}}$ for $\xi > 5\times 10^{-6}$ is calculated by averaging the results obtained  from Eq.~(\ref{eq:AccurateCrossSection}) for the  different pileup conditions of Table~\ref{pileupcorr}.
The largest systematic uncertainty, besides the 4\% uncertainty of the absolute luminosity value,  is due to fluctuations in the luminosity determination of the different low-pileup runs. The model dependence of the efficiency $\epsilon_\xi$ contributes $\pm$1\%, while the correction for the contamination from  events below the $\xi$ threshold is uncertain by $\pm$0.5\%  as given by the standard deviation of the $(1-b_{\xi})$ factors obtained from the three MC simulations studied.  The exclusion of  noisy HF towers in the calculation of HF energy changes the results by  $\pm$0.4\%, a value that is taken as a systematic uncertainty. Finally, lowering  the value of the calorimeter threshold  $E_{\mathrm{HF}}$ from 5 to 4\GeV introduces a change of 0.2\% in the final result.

\begin{table*}[htb]
  \begin{center}
\topcaption{List of systematic sources and their effects on the value of the inelastic cross section measured using HF calorimeters. The integrated luminosity contributes an additional uncertainty of 4\% to this measurement.}
  \begin{tabular}{l  c c } \hline
Systematic source                     &  Uncertainty on $\sigma_{\text{inel}}$  & Change in $\sigma_{\text{inel}}$ \\ \hline
\text{Run-to-run variation}           & $\pm$0.8\unit{mb}       &  $\pm$1.3\%   \\
\text{Selection efficiency}           & $\pm$0.6\unit{mb}       &  $\pm$1.0\%  \\
\text{Contamination from $\xi < 5\times 10^{-6}$ } &$\pm$0.3\unit{mb} & $\pm$0.5\%\\
\text{HF tower exclusion }            & $\pm$0.3\unit{mb}       &  $\pm$0.4\% \\
\text{HF energy threshold }           & $\pm$0.1\unit{mb}       &  $\pm$0.2\% \\
\hline
\text{Total} (in quadrature)          & $\pm$1.1\unit{mb}  & $\pm$1.8\%\\
\hline
\end{tabular}
\label{tab:HFmethsyst}
\end{center}
\end{table*}

Table~\ref{tab:HFmethsyst}  lists the individual systematic uncertainties,  and
their total impact, calculated by adding the separate contributions in quadrature. The inelastic pp cross section for  events with $\xi > 5\times 10^{-6}$ is found to be:
\begin{equation}\begin{split}
\sigma_{\text{inel}}(\xi > 5\times 10^{-6}) &= [ 60.2 \pm 0.2\stat \pm 1.1\syst\\
&\qquad \pm 2.4\lum]\unit{mb.}
\end{split}\end{equation}

This result is in agreement  with equivalent measurements from the ATLAS Collaboration $\sigma^{\text{ATLAS}}_{\text{inel}}(\xi > 5\times 10^{-6}) = [ 60.3 \pm 0.05\stat\pm0.5\syst\pm 2.1\lum]$\unit{mb}~\cite{ATLASXSEC}, and from the ALICE Collaboration $\sigma^{\text{ALICE}}_{\text{inel}}(\xi > 5\times 10^{-6}) = [ 62.1^{~1.0}_{-0.9}\syst\pm 2.2\lum]$\unit{mb}~\cite{:2012sj}. The uncertainties on luminosity of the three measurements  are highly correlated.

\section{Estimating the inelastic  cross section by counting event vertices}
\label{TRKMethod}

A vertex-counting method is also used to measure the inelastic pp cross section. The method  relies  on the accuracy of the CMS tracking system and  not upon any specific Monte Carlo simulation.  This method assumes that the number ($n$) of inelastic pp interactions in a given bunch crossing follows  the Poisson probability distribution of Eq.~(\ref{eq:Poisson}), where $\lambda$ is calculated from the product of the  instantaneous luminosity  for a bunch crossing and the total inelastic pp cross section: $\lambda = L\cdot \sigma_{\text{inel}}$. The probability of having $n$ inelastic pp interactions, each producing a vertex  with  $>$1, $>$2, or $>$3 charged particles with $\pt > 200 \MeVc$ within $|\eta| = 2.4$, for $n$ between 0 and 8,  is measured   at different  luminosities to evaluate  $\sigma_\text{inel}$ from a  fit  of Eq.~(\ref{eq:Poisson}) to the data.

\subsection{Event selection and method of  analysis}

Inclusive samples of ${\approx}3\times 10^6$ two-electron candidate events, and ${\approx}1.5\times 10^6$ single-muon candidate events, are selected for this analysis. The specific trigger requirements  are not important, as long as their efficiencies do not depend on the number of pileup interactions. The ``triggering interaction'', i.e. the process associated with the production of either the two electrons  or the single muon,  is not included in the vertex count, but is used just to sample unbiased pileup interactions, given by the additional vertices in the same bunch crossing.

The analysis is performed using data collected with the single-muon sample, while the data collected with the two-electron trigger are used to perform  a systematic check on the effect of the choice of the trigger on the result.  For each of the two data samples, the distributions in the fraction of events with 0  to 8 pileup interactions are measured as a function of luminosity. A bin-by-bin  correction is applied to these measurements to obtain  true distributions which are then fitted to Eq.~(\ref{eq:Poisson}), to extract a common value of $\sigma_\text{inel}$. This correction is mainly due to vertex reconstruction efficiency and $\pt$ migration. The distribution of the bin-by-bin  correction factors  is centered around 1  with all values contained in the interval 0.7--1.3.

The bin-by-bin corrections, evaluated from full Monte Carlo simulation (\PYTHIA6) and reconstruction of events in the CMS detector, do not depend on any specific production model, but only on an accurate simulation of the CMS tracking system.  The distributions  of charged particles in transverse momentum   and in track multiplicity  in MC events are reweighted to provide  agreement with the data, as these two quantities influence the vertex reconstruction efficiency. The track multiplicity distribution has a broad maximum between 4 and 8 and extends up to 70 tracks.   Cross sections are measured for inclusive pp interactions with $>$1, $>$2, and $>$3 charged particles, with $\pt > 200$\MeVc and $|\eta| < 2.4$, where ``charged particles'' refer to those  with decay lengths $c\tau > 1\unit{cm}$.

\subsection{Vertex  definition and reconstruction}
\label{sub:vertex}

To be counted, a pileup interaction has to have a sufficient number of tracks to provide a vertex of good quality~\cite{:CMSTRK}. The vertex quality depends upon the number and  characteristics of the individual tracks attributed to each vertex. A vertex is also required to have the longitudinal $z$ position within 20\unit{cm} of the nominal interaction point.

There are two main reasons that lead to incorrect vertex  reconstruction: (i) overlap with another vertex, i.e. the reconstruction program merges two vertices, and (ii) an insufficient number of tracks, or  tracks  too poorly measured to  pass vertex-quality requirements. The  vertexing algorithm is very efficient in distinguishing vertices that are further apart than 0.06\unit{cm} along the beam direction, and this analysis requires a minimum distance of 0.1 cm. Minimum distances of 0.06\unit{cm} and of 0.2\unit{cm} are used to check for any systematic effects from this requirement. The fraction of vertices lost from merging depends on  luminosity and  is almost negligible in the lowest luminosity bin while it becomes around  2\% in the highest bin, an effect that is well reproduced by MC  simulation, and is therefore corrected. The second source of inefficiency in vertex reconstruction depends on the number of  tracks per vertex. Vertices with a large number of tracks are always well reconstructed, while vertices with less than 10 tracks suffer some degradation in reconstruction efficiency: this efficiency is 80\% for four-track  vertices, 65\%  for three-track vertices and 40\% for two-track vertices.

There are also two main sources of  secondary vertices that are not  related specifically to  $\sigma_{\text{inel}}$: additional vertices generated through decays of long-lived particles, and  false secondary vertices generated by splitting a single vertex into two distinct vertices. Misidentified secondary vertices can often be  rejected, as they have a much lower track multiplicity, and they are not necessarily positioned along the beam line: for this last reason, the transverse position of the vertex is required to be within $\pm 0.06$\unit{cm} from the nominal beam line.

The correction of number of  candidate vertices to the true  number of pileup interactions is considered as a function of luminosity. In particular, a 2-vertex event recorded at low luminosity is most likely to correspond to a true 2-vertex event, while a 2-vertex event recorded at high luminosity is most likely a 3 or 4-vertex event, in which 1 or 2 vertices are merged. We divide the data into 13 equal intervals of  instantaneous bunch-crossing luminosity, from $0.05 \times 10^{30}$ to $0.7 \times 10^{30}\unit{cm}^{-2} \unit{s}^{-1}$. To obtain the true pileup  distribution in each luminosity interval, we proceed as follows:
\newline
 (i) Using Eq.~(\ref{eq:Poisson}), the expected distribution of pileup interactions is calculated  for the specific luminosity interval, assuming some trial value $\sigma^{\text{trial}}_{\text{inel}}$ for the inelastic cross section.
\newline
 (ii) The Monte Carlo simulation is reweighted to  generate a pileup  distribution matching the one calculated in step (i).  Steps (i) and (ii) are repeated several times for different   $\sigma^{\text{trial}}_{\text{inel}}$, until good agreement is reached between data and the reconstructed pileup distributions for MC events.
\newline
 (iii) The generated  pileup distributions for inclusive interactions with $>$1, $>$2, and $>$3 tracks, each with $\pt > 200$\MeVc  and $|\eta| < 2.4$, is obtained from the reweighted Monte Carlo.
\newline
 (iv) The bin-by-bin corrections are computed using the ratio of reconstructed to generated Monte Carlo pileup distributions  for $>$1, $>$2,  and $>$3 tracks, yielding thereby the correction factors for each of these three inclusive sets of events.

\begin{figure*}[htb]
\begin{center}
  \includegraphics[width=\linewidth]{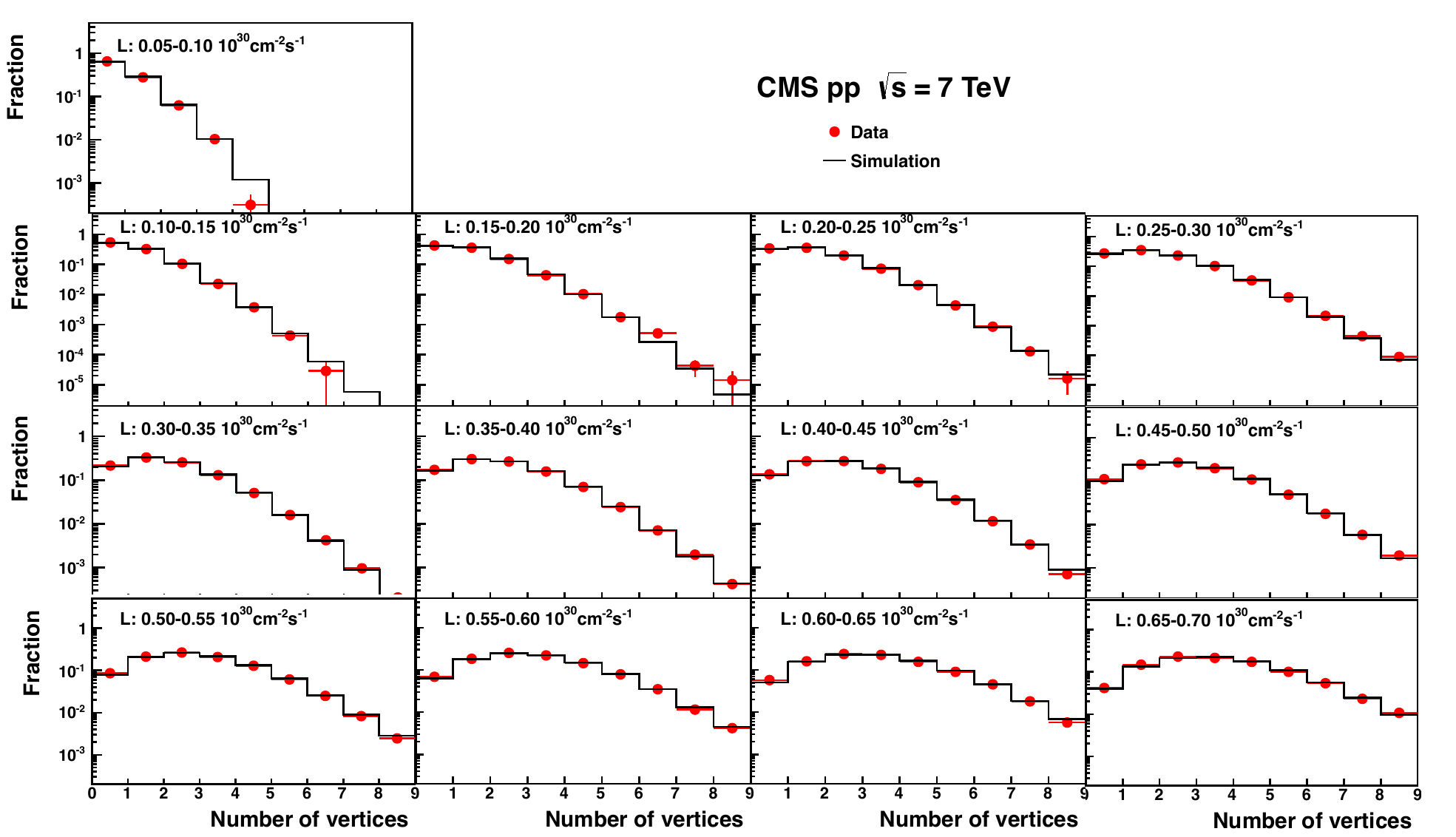}
 \end{center}
 \caption{Fraction of reconstructed  events  with more than one track, corrected for efficiency, measured as a function of the number of vertices, in data (dots) and in Monte Carlo (histogram), for instantaneous bunch-crossing  luminosities between $0.05\times 10^{30}$ and $0.7\times 10^{30}\percms$.}
 \label{fig:MC_data}
 \end{figure*}

The corrected fractional distributions of events, for interactions with more than 1 track in data or in the MC,  are compared in  Fig.~\ref{fig:MC_data} as a function of the number of vertices ($n$) for the thirteen bins in instantaneous luminosity.

\subsection{Results and systematic uncertainties}

\begin{figure*}[htb]
\begin{center}
 \includegraphics[width=\linewidth]{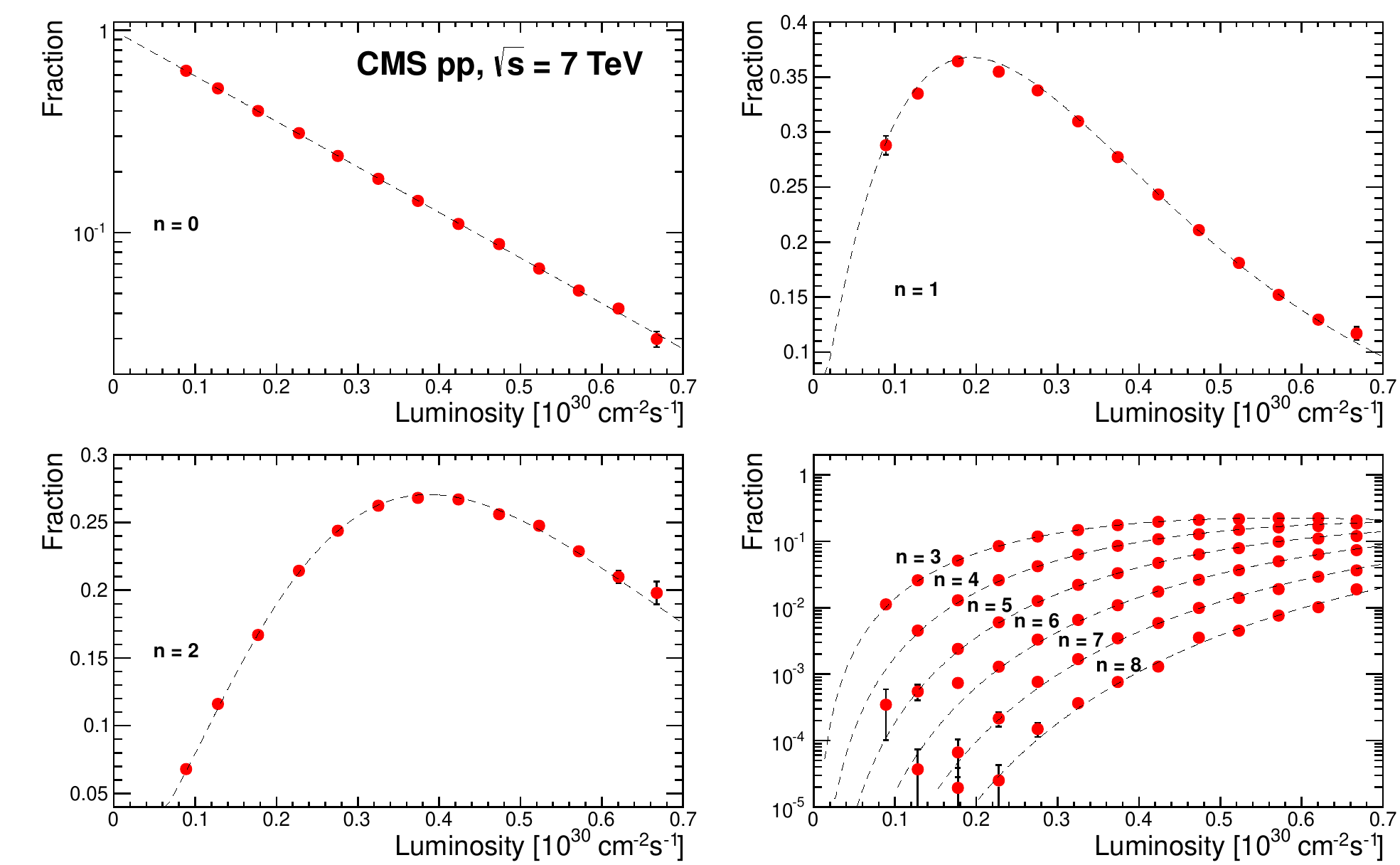}
 \end{center}
 \caption{Fraction of pp events with $n$ pileup  vertices, for $n = 0$ to 8, containing more than one charged particle, as a function of instantaneous  bunch-crossing luminosity. The dashed lines are  the fits described in the text. The data points are plotted at the mean of the differential distribution in each bin.}
 \label{fig:final}
 \end{figure*}

\begin{figure}[htb]
\begin{center}
 \includegraphics[width=\linewidth]{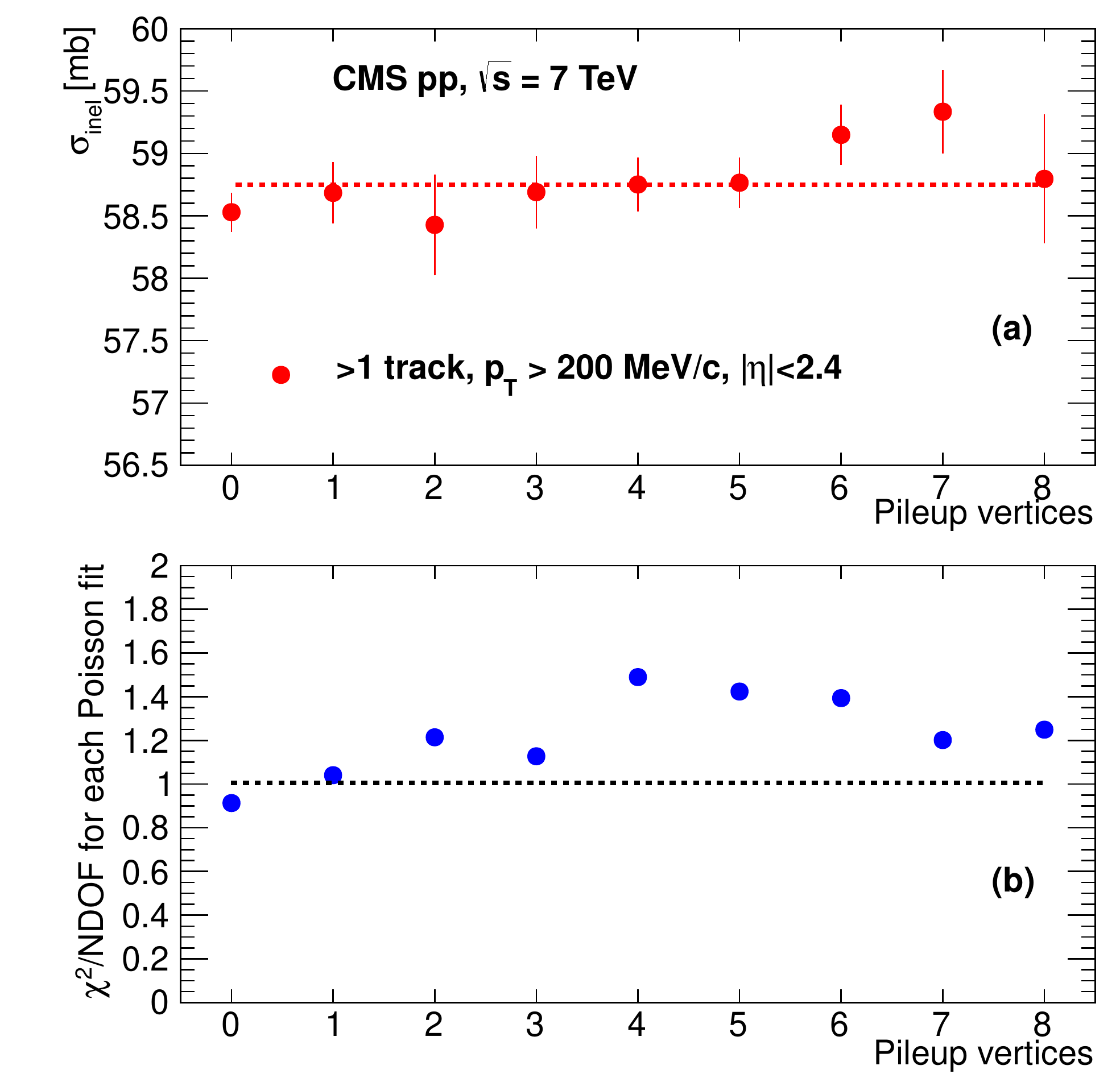}
 \end{center}
\caption{(a) Values of the inelastic pp cross section  $\sigma_\text{inel}$ and   (b) their associated goodness of fit $\chi^2/\mathrm{NDOF}$, obtained for each of the fits in Fig.~\ref{fig:final}, as a function of the number of pileup vertices, in interactions with $>1$ track with $\pt > 200$\MeVc and $|\eta| < 2.4$. The  line in (a) is the result of a fit to the 9 individual values of $\sigma_\text{inel}$, while the dashed line in (b) indicates  $\chi^2/\mathrm{NDOF} = 1$.}
 \label{fig:sigma}
 \end{figure}

Figure \ref{fig:final} displays the data points from Fig.~\ref{fig:MC_data} as a function of the instantaneous luminosity, for events with $n$ = 0 to 8 pileup vertices. For each  $n$, the values of the Poisson distribution given by  Eq.~(\ref{eq:Poisson})  are fitted as a function of $\lambda = L \cdot \sigma_\text{inel}$ to the data, providing
nine estimates of the inelastic cross section. Their weighted average  provides  the final result shown in Fig.~\ref{fig:sigma} (a).  The error bars and the values of goodness of fit per degree of freedom ($\chi^2/\mathrm{NDOF}$) for each result are obtained from the individual Poisson fits of Fig.~\ref{fig:final}. Figure \ref{fig:sigma} (b) shows the normalized $\chi^2$ values for each of the fits to Eq.~(\ref{eq:Poisson}). The equivalent plots for vertices with more than two and more than three tracks are very similar, as the overlap among the datasets is above 95\%.

The main source of systematic uncertainty is the 4\% uncertainty on  CMS luminosity, which leads to an uncertainty $ \Delta\sigma_{\text{lum}} = \pm2.4\unit{mb}$.  The  largest contribution arising from  the method of analysis,  $\Delta\sigma_{\mathrm{vtx}} = \pm1.4\unit{mb}$, is  the uncertainty on the vertex-reconstruction efficiency,  which is evaluated using a Monte Carlo simulation   and a method based on data. This second technique utilizes measured quantities such as the distribution of the longitudinal $z$ position of the vertex and the distribution of the minimum distance between two vertices to evaluate the vertex-reconstruction efficiency. Other uncertainties  linked to vertex selection are estimated by: (i) reducing the range used for accepting longitudinal positions of vertices from $|z|<20$ to $|z|<$~10\unit{cm} relative to the centre of the CMS detector, (ii)  modifying the vertex-quality requirements, (iii) changing the minimum distance between two vertices from $\Delta z <  0.1\unit{cm}$ to 0.06\unit{cm} and 0.2\unit{cm}, and (iv) changing the maximum allowed transverse coordinate of the vertex from ${\pm}0.06\unit{cm}$ to ${\pm}0.05\unit{cm}$ and ${\pm}0.08\unit{cm}$.

Several other possible sources of uncertainty have also been checked by: (i) performing the analysis  on sets of data collected with different trigger requirements (two-electron or single-muon trigger) to measure the effect of the trigger on the selection of pileup events, (ii) changing the luminosity interval used in the fit by ${\pm}0.05 \times 10^{30}\unit{cm}^{-2}\unit{s}^{-1}$,
and (iii) repeating the analysis without reweighting the  track-multiplicity distributions in the MC, %ROBI
 to evaluate the effect of an incorrect track-multiplicity shape, which should not influence the bin-by-bin correction to first order.
The uncertainty attributed to each systematic source is defined by the largest change in  $\sigma_\text{inel}$.
The full list of the systematic sources is shown in Table~\ref{tab:PUmethsyst}. Adding all the uncertainties in quadrature yields a  total systematic uncertainty on the method of $\Delta\sigma_\text{syst} = {\pm}2.0\unit{mb}$.

The measured values of $\sigma_\text{inel}$ for inclusive interactions with $>$1, $>$2, and $>$3 charged particles with $|\eta| < 2.4$ and  $\pt > 200$\MeVc, as well as their individual uncertainties, are listed in Table~\ref{tab:PUmethresults}. The statistical error is below 0.1\unit{mb} and is ignored.

\begin{table*}[htb]
  \begin{center}
\topcaption{List of systematic sources and their effects on the value of the inelastic cross section measured  using  the vertex-counting method. The \% changes are shown for the results of the $\sigma_{\text{inel}}({>}1\unit{track})$ measurement.  The integrated luminosity contributes an additional uncertainty of 4\% to this measurement.}
  \begin{tabular}{l  c c } \hline
Systematic source   &  Uncertainty on $\sigma_{\text{inel}}$  & Change in $\sigma_{\text{inel}}({>}1\unit{track})$  \\ \hline
Vertex reconstruction efficiency     & $\pm$1.4\unit{mb}  &  $\pm$2.4\%\\
Longitudinal position of vertex      & $\pm$0.1\unit{mb}  &  $\pm$0.2\% \\
Vertex quality                   & $\pm$0.7\unit{mb}  &  $\pm$1.3\% \\
Minimum distance between vertices   & $\pm$0.1\unit{mb}  &  $\pm$0.2\% \\
Transverse position of vertex    & $\pm$0.3\unit{mb}  &  $\pm$0.6\% \\
Different sets of data           & $\pm$0.9\unit{mb}         &  $\pm$1.6\%   \\
Range of luminosity used in fit     & $\pm$0.2\unit{mb}   &  $\pm$0.4\% \\
Reweighting MC track distribution    & $\pm$0.2\unit{mb}   &  $\pm$0.4\% \\
\hline
\text{Total (in quadrature)}   & $\pm$2.0\unit{mb}  &  $\pm$3.3\% \\ \hline

\end{tabular}
\label{tab:PUmethsyst}
\end{center}
\end{table*}

\begin{table}[h]
  \begin{center}
\topcaption{$\sigma_\text{inel}$ values for interactions with $>$1, $>$2 and $>$3 charged particles, with their uncertainties from systematic sources of the method and from luminosity. The statistical error is below 0.1\unit{mb} and is ignored.}
  \begin{tabular}{c  c } \hline
 Measurement  &  Result   \\ \hline
 $\sigma_\text{inel}({>}1\unit{track})$ & $[ 58.7 \pm 2.0\syst\pm 2.4\lum]$\unit{mb}  \\
 $\sigma_\text{inel}({>}2\unit{tracks})$ & $[ 57.2 \pm 2.0\syst\pm 2.4\lum]$\unit{mb}  \\
 $\sigma_\text{inel}({>}3\unit{tracks})$ & $[ 55.4 \pm 2.0\syst\pm 2.4\lum]$\unit{mb}  \\
\hline
\end{tabular}
\label{tab:PUmethresults}
\end{center}
\end{table}

\section{Results and comparison with Monte Carlo models}
\label{Results}

The two  techniques presented to measure the
inelastic pp cross section complement each other. The calorimeter-based method is very
sensitive to events that produce forward energy deposition, and, in particular, small $M_X$ values that comprise particle systems highly
boosted along the beam line. However, the method is less sensitive to central diffractive dissociation events, with particle production concentrated at  small pseudorapidities. Conversely, the vertex-counting method is geared toward  measurement of
centrally-produced events, and is not optimal for events with particles produced mostly at large $\eta$.
The concurrent use of these two methods provides therefore  almost complete coverage of all types of pp inelastic events, with particle production in the range of $ \abs{\eta} \lesssim 5$.

\begin{figure}[htb]
\begin{center}
  \includegraphics[width=\cmsFigWidth]{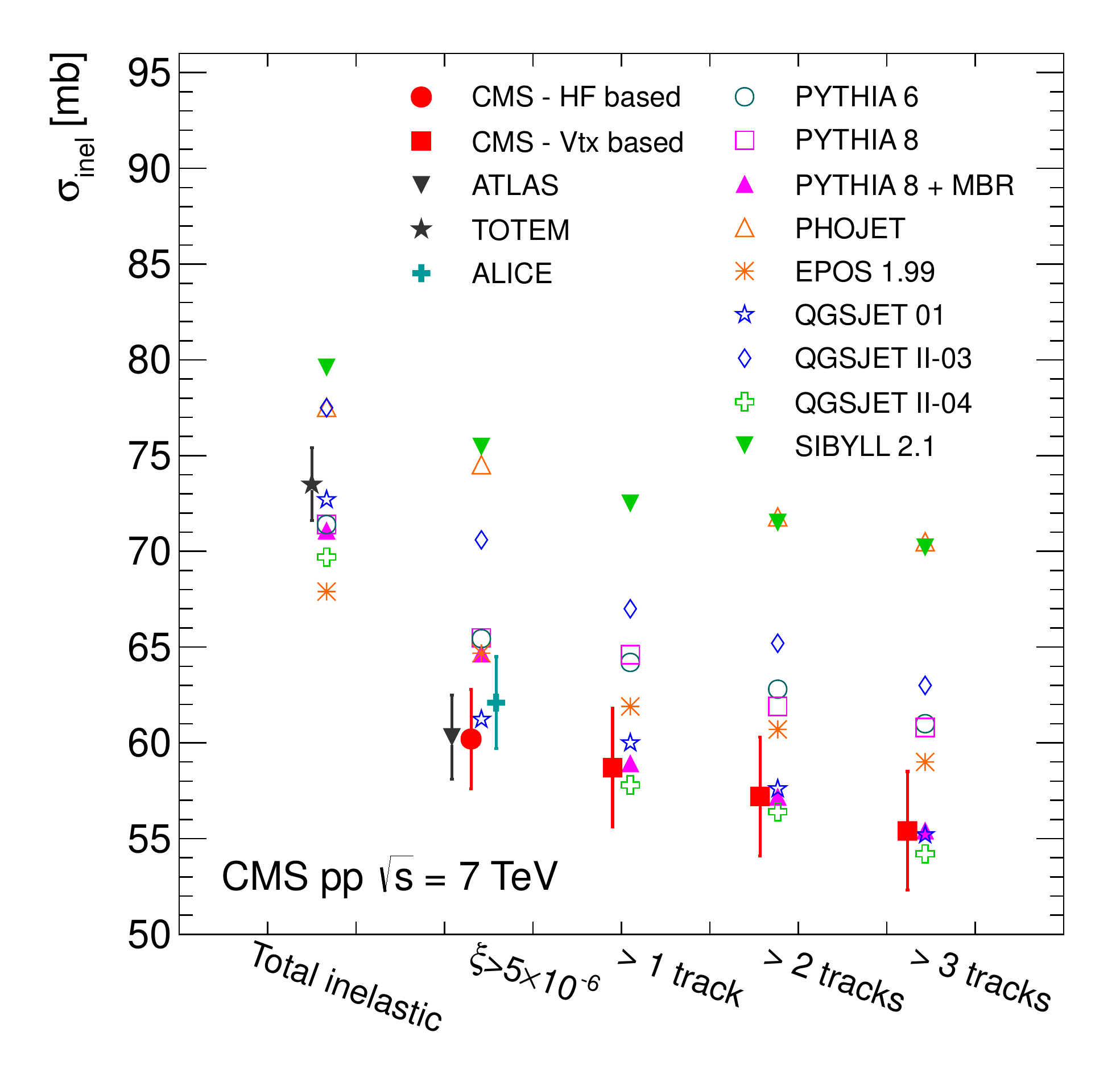}
 \end{center}
 \caption{The two types of CMS measurements of the inelastic pp cross section (red filled circle and squares) compared to  predictions from several Monte Carlo models for different criteria, as labelled below the abscissa axis. The MC predictions have an uncertainty of 1\unit{mb} (not shown). The label \PYTHIA6 (tunes D6T, Z1\_LEP, AMBT1, DW-Pro, and
Pro-PT0) and \PYTHIA8 (versions  8.127--8.139, Tunes 2C 8.140 Cor10a, Tune 2M 8.140 Cor10a, and Tune 4C 8.145 Cor10a)  indicates several versions that give equivalent results. Other LHC experimental results are also included for comparison.}
 \label{fig:sigma_comp}
 \end{figure}

Figure~\ref{fig:sigma_comp}   compares the CMS results with the measurements presented by the
TOTEM~\cite{Antchev:2011vsR}, the ATLAS~\cite{ATLASXSEC} and the ALICE~\cite{:2012sj} collaborations, as well as with predictions of
two groups of Monte Carlo models. The first group comprises several versions of \PYTHIA:
\PYTHIA6 (tunes D6T, Z1\_LEP~\cite{Field:2010bc}, AMBT1, DW-Pro, and
Pro-PT0 provide very similar results), \PYTHIA8
(versions  8.135 Tune 1, 8.145 Tunes 2C, Tune 2M, and Tune 4C  are equivalent) and the recent \PYTHIA8 MBR tune~\cite{Ciesielski:2012mc} (version 8.165). The second group includes MC generators
based on the  same Regge-Gribov phenomenology, but with different implementations of model
ingredients~\cite{d'Enterria:2011kwR}:  \textsc{phojet}, as well as three MC
programs commonly used in cosmic-rays physics,  such as \textsc{qgsjet}~01~\cite{Ostapchenko:2004ss},
\textsc{qgsjet}~II (versions 03 and 04)~\cite{Ostapchenko:2010vb},
\textsc{sibyll} (version 2.1)~\cite{Fletcher:1994bd} and
\textsc{epos} (version 1.99)~\cite{Werner:2005jf}.

The \textsc{phojet} and \textsc{sibyll} models overestimate the observed cross sections by more than 20\%, while the \textsc{epos}, \textsc{qgsjet}~II-03, \PYTHIA6, and \PYTHIA8 tunes provide predictions that are about 10\% larger than the measured inelastic cross sections. \textsc{qgsjet 01} and \textsc{qgsjet}~II-04  agree within one  standard deviation with the data points. The \PYTHIA8-MBR tune reproduces rather well the vertex-based measurements, while it overestimates the calorimeter-based result.

\begin{figure}[htb]
\begin{center}
  \includegraphics[width=\cmsFigWidth]{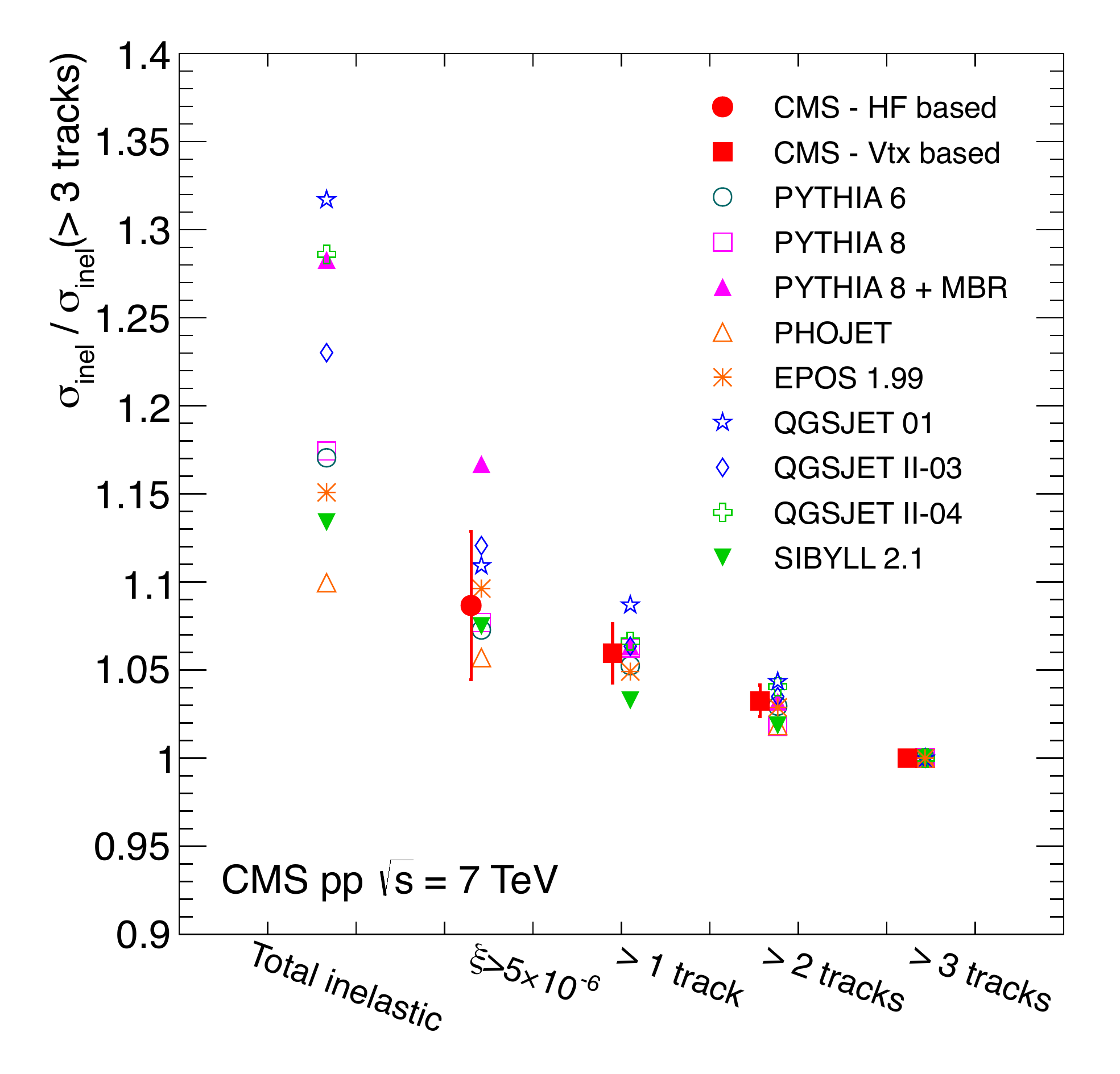}
 \end{center}
 \caption{Comparison of the measured inelastic pp cross sections with  predictions  of several Monte Carlo models, for different criteria, normalized to the value obtained for $>$3 tracks.}
 \label{fig:sigma_norm}
 \end{figure}

A comparison of the trends in the data with the MC models is shown in Fig.~\ref{fig:sigma_norm}, where the cross sections are now normalized to the $\sigma_\text{inel}$  value measured for events with $>$3 tracks.  In these ratios both the systematic and statistical uncertainties are reduced as the correlations between the four measurements are very large. The values and uncertainties of the cross sections ratios are shown in Table~\ref{tab:xsecratios}.
The dependence of $\sigma_\text{inel}$  on the nature of the final states relative to the results for $>$3 tracks, is  well reproduced  by most MC simulations.

\begin{table}[h]
  \begin{center}
\topcaption{Measured inelastic pp cross sections normalized to $\sigma_\text{inel}({>}3\unit{tracks})$, and their uncertainties.}
  \begin{tabular}{l  c } \hline
 Ratio  &  Result   \\ \hline
 $\sigma_\text{inel}({>}2\unit{tracks})$/$\sigma_\text{inel}({>}3\unit{tracks})$ & $ 1.032\pm 0.009$  \\
 $\sigma_\text{inel}({>}1\unit{track})$/$\sigma_\text{inel}({>}3\unit{tracks})$  & $ 1.060 \pm 0.017$  \\
 $\sigma_\text{inel}(\xi > 5\times 10^{-6})$/$\sigma_\text{inel}({>}3\unit{tracks})$ & $ 1.087 \pm 0.042$ \\
\hline
\end{tabular}
\label{tab:xsecratios}
\end{center}
\end{table}

The TOTEM collaboration~\cite{Antchev:2011vsR} has recently measured a total pp inelastic cross section of $\sigma_{\text{inel}} = 73.5^{+2.4}_{-1.9}$\unit{mb}.  Although several Monte Carlo models such as \textsc{epos}, \textsc{qgsjet 01}, \textsc{qgsjet}~II-4,  \PYTHIA6, and  \PYTHIA8 reproduce this value  (Fig.~\ref{fig:sigma_comp}), only \textsc{qgsjet}~01  and  \textsc{qgsjet}~II-04, and  \PYTHIA8-MBR (but less so)  are able  to simultaneously reproduce  the  less inclusive CMS measurements. This observation suggests that most of the  Monte Carlo models overestimate the contribution from high-mass diffraction to the total inelastic cross section, and underestimate the component at low mass.

\section{Summary}
\label{Conclusions}

The inelastic cross section in pp collisions at $\sqrt{s} = 7\TeV$ has been measured using two  methods that incorporate information either from central or from  forward detectors of CMS.
The results for the different choices of final states considered are:
\begin{align*}
&\sigma_\text{inel}(\xi > 5\times 10^{-6}) = [ 60.2 \pm 0.2\stat\pm 1.1\syst\\
&\qquad\pm 2.4\lum]\unit{mb},\\
&\sigma_\text{inel}(>1\unit{track}) = [ 58.7 \pm 2.0\syst\pm 2.4\lum]\unit{mb},\\
&\sigma_\text{inel}(>2\unit{tracks})  = [ 57.2 \pm 2.0\syst\pm 2.4\lum]\unit{mb},\\
&\sigma_\text{inel}(>3\unit{tracks}) = [ 55.4 \pm 2.0\syst\pm 2.4\lum]\unit{mb},\\
\end{align*}
where each track must have $\pt > 200$\MeVc and $|\eta| < 2.4$. The comparison of  these results with the cross  section expected from Monte Carlo models used in collider and cosmic-rays studies shows  that \textsc{phojet} and \textsc{sibyll} largely  overestimate  $ \sigma_\text{inel}$. The \textsc{epos},  \textsc{qgsjet}~II-03, \PYTHIA6, and \PYTHIA8 (except the MBR tune) programs predict values about 10\% above the data, while  \textsc{qgsjet 01}, \textsc{qgsjet}  II-04 agree well with the measurements. \PYTHIA8+MBR agrees well with the track-based measurements, but overestimates the prediction  for $\sigma_\text{inel}$ for $\xi > 5\times 10^{-6} $.  All models agree broadly with the relative dependence of the cross section on the criteria used to define the final states.

\section*{Acknowledgments}

We congratulate our colleagues in the CERN accelerator departments for the excellent performance of the LHC machine. We thank the technical and administrative staffs at CERN and other CMS institutes, and acknowledge support from BMWF and FWF (Austria); FNRS and FWO (Belgium); CNPq, CAPES, FAPERJ, and FAPESP (Brazil); MEYS (Bulgaria); CERN; CAS, MoST, and NSFC (China); COLCIENCIAS (Colombia); MSES (Croatia); RPF (Cyprus); MoER, SF0690030s09 and ERDF (Estonia); Academy of Finland, MEC, and HIP (Finland); CEA and CNRS/IN2P3 (France); BMBF, DFG, and HGF (Germany); GSRT (Greece); OTKA and NKTH (Hungary); DAE and DST (India); IPM (Iran); SFI (Ireland); INFN (Italy); NRF and WCU (Korea); LAS (Lithuania); CINVESTAV, CONACYT, SEP, and UASLP-FAI (Mexico); MSI (New Zealand); PAEC (Pakistan); MSHE and NSC (Poland); FCT (Portugal); JINR (Armenia, Belarus, Georgia, Ukraine, Uzbekistan); MON, RosAtom, RAS and RFBR (Russia); MSTD (Serbia); SEIDI and CPAN (Spain); Swiss Funding Agencies (Switzerland); NSC (Taipei); ThEP, IPST and NECTEC (Thailand); TUBITAK and TAEK (Turkey); NASU (Ukraine); STFC (United Kingdom); DOE and NSF (USA).

\bibliography{auto_generated}   % will be created by the tdr script.

\cleardoublepage \appendix\section{The CMS Collaboration \label{app:collab}}\begin{sloppypar}\hyphenpenalty=5000\widowpenalty=500\clubpenalty=5000\textbf{Yerevan Physics Institute,  Yerevan,  Armenia}\\*[0pt]
S.~Chatrchyan, V.~Khachatryan, A.M.~Sirunyan, A.~Tumasyan
\vskip\cmsinstskip
\textbf{Institut f\"{u}r Hochenergiephysik der OeAW,  Wien,  Austria}\\*[0pt]
W.~Adam, E.~Aguilo, T.~Bergauer, M.~Dragicevic, J.~Er\"{o}, C.~Fabjan\cmsAuthorMark{1}, M.~Friedl, R.~Fr\"{u}hwirth\cmsAuthorMark{1}, V.M.~Ghete, J.~Hammer, N.~H\"{o}rmann, J.~Hrubec, M.~Jeitler\cmsAuthorMark{1}, W.~Kiesenhofer, V.~Kn\"{u}nz, M.~Krammer\cmsAuthorMark{1}, I.~Kr\"{a}tschmer, D.~Liko, I.~Mikulec, M.~Pernicka$^{\textrm{\dag}}$, B.~Rahbaran, C.~Rohringer, H.~Rohringer, R.~Sch\"{o}fbeck, J.~Strauss, A.~Taurok, W.~Waltenberger, G.~Walzel, E.~Widl, C.-E.~Wulz\cmsAuthorMark{1}
% ...
\vskip\cmsinstskip
\dag:~Deceased\\
1:~~Also at Vienna University of Technology, Vienna, Austria\\
% ...

\end{sloppypar}
\end{document}